\documentclass[a4paper,11pt]{article}
\pdfoutput=1
\usepackage{jheppub}

\usepackage[usenames,dvipsnames]{xcolor}
\usepackage{indentfirst}
\usepackage{amsmath}
\usepackage{paralist}
\usepackage{bbold}

\usepackage{tikz}
\usetikzlibrary{intersections}
\usetikzlibrary{arrows,decorations.pathmorphing,backgrounds,positioning,fit,petri}
\usetikzlibrary{decorations.markings}
\usetikzlibrary{calc}
\usetikzlibrary{intersections,through,hobby}
\usetikzlibrary{external}

\pgfdeclarelayer{edgelayer}
\pgfdeclarelayer{nodelayer}
\pgfsetlayers{edgelayer,nodelayer}

\usepackage{cleveref}

\definecolor{colA}{HTML}{c19277}
\definecolor{colB}{HTML}{e1bc91}
\definecolor{colD}{HTML}{62959c}

\hypersetup{
  citecolor=colA, 		
  linkcolor=colA,    	
  urlcolor=colD,			
}

\newcommand{\ep}{\ensuremath{\varepsilon}}
\newcommand{\zb}{\ensuremath{\bar{z}}}
\newcommand{\MS}{\ensuremath{\overline{\rm MS}}}
\newcommand{\sru}{\ensuremath{e^{\frac{i\pi}{3}}}}

\newcommand{\icap}[2] {\texttt{\tiny #1[#2]}}

\title{Three-loop vertex integrals at symmetric point  }

\author[a,b]{Andrey Pikelner}
\affiliation[a]{Bogoliubov Laboratory of Theoretical Physics,
  Joint Institute for Nuclear Research,\\
  Dubna 141980, Russia}
\affiliation[b]{Theory department, Budker Institute of Nuclear Physics, \\
  Novosibirsk 630090, Russia}

\emailAdd{pikelner@theor.jinr.ru}

\abstract{
  This paper provides details of the massless three-loop three-point integrals
  calculation at the symmetric point. Our work aimed to extend known two-loop
  results for such integrals to the three-loop level.
  Obtained results can find their application in regularization-invariant
  symmetric point momentum-subtraction (RI/SMOM) scheme QCD calculations of
  renormalization group functions and various composite operator matrix elements.
  To calculate integrals, we solve differential equations for auxiliary
  integrals by transforming the system to the $\varepsilon$-form.
  Calculated integrals are expressed through the basis of functions with uniform
  transcendental weight. We provide expansion up to the transcendental weight six
  for the basis functions in terms of harmonic polylogarithms with six-root of
  unity argument.
}

\begin{document}
\maketitle


\section{Introduction}
\label{sec:intro}
Many physically significant quantities in QCD can be extracted from the
three-point Green functions. There is remarkable
progress~\cite{Baikov:2016tgj,Luthe:2017ttg,Herzog:2017ohr,Chetyrkin:2017bjc} in
calculating QCD renormalization group functions in the minimal
subtraction($\MS$) renormalization scheme~\cite{tHooft:1973mfk}. We can choose
exceptional momenta routing(one of the external momenta set to zero) and
restrict ourselves to considering only the divergent parts of diagrams for these
calculations. Due to the unphysical nature of the minimal subtraction scheme, a
group of momentum subtraction schemes(MOM)~\cite{Celmaster:1979km} is widely
used in calculations requairing finite parts of the three-point functions. Such
physical schemes are crucial for the Lattice calculations, where one has access
to vertex functions directly. Calculation of various vertices in regularization
invariant momentum subtraction (RI/MOM) schemes provides a connection between
the lattice calculations and the MS scheme results. By choosing exceptional
momenta routing, putting one of the external momenta to zero, we define RI/MOM
scheme. With symmetric point kinematics configuration with all squares of
external momenta equal, we define RI/SMOM scheme.

The main difficulty in the RI/MOM and RI/SMOM scheme calculation is the
necessity to know finite parts of vertex functions in the chosen kinematics.
If we require only finite parts of the propagator type integrals for the exceptional
momenta routing, then for the vertices in the RI/SMOM case, we need to know
three-point integrals at the symmetric point.
Calculation of the symmetric point integrals is the main difficulty of the
RI/SMOM scheme. The present paper focuses on how to solve the problem at the
three-loop level.

There are many results based on the calculation of the two-loop three-point
functions at the symmetric point. Namely three-loop RI/SMOM
beta-functions\cite{Gracey:2011pf}, two-loop correction to the relation between
RI/SMOM and MS quark mass\cite{Almeida:2010ns,Gracey:2011fb}, and
renormalization of Wilson operator matrix
elements\cite{Almeida:2010ns,Gracey:2011pf,Gracey:2011fb,Gracey:2011zn,Gracey:2011zg}.
We recently extended some of these calculations to the three-loop level
analytically\cite{Bednyakov:2020ugu,Bednyakov:2020cdf}, using results of the
present paper. Independently, numerical results for renormalization of
different operators matrix elements appeared in the series of
papers\cite{Kniehl:2020nhw,Kniehl:2020sgo}.

Integrals of our interest are given on the left of the Fig.~\ref{fig:kin3pt}
and condition $p_1^2 = p_2^2 = q^2 = -1$ defines the symmetric subtraction
point.
\tikzset{
  di/.style={line width=1pt,draw=black, postaction={decorate},
    decoration={markings,mark=at position .65 with
      {\arrow[scale=1,draw=black,>=latex]{>}}}},
  diA/.style={line width=1pt,draw=colA, postaction={decorate},
    decoration={markings,mark=at position .65 with
      {\arrow[scale=1,draw=colA,>=latex]{>}}}},
  ndi/.style={line width=1pt,draw=black}
}
\begin{figure}[h]
  \centering
  \begin{tikzpicture}[baseline={(current bounding box.center)}]
    \useasboundingbox (-2,-2) rectangle (2,2);
    \begin{pgfonlayer}{edgelayer}
      \filldraw[fill=colD!50!white, draw=black, line width=1pt] (0,0) circle (0.8cm);
      \draw[di] (90:1.6) -- (90:0.8);
      \draw[di] (215:0.8) -- (215:1.6);
      \draw[di] (325:0.8) -- (325:1.6);
      \draw (90:1.5) node[anchor=west] {$q$};
      \draw (215:1.5) node[anchor=south east] {$p_1$};
      \draw (325:1.5) node[anchor=south west] {$p_2$};
    \end{pgfonlayer}
  \end{tikzpicture}
  \hspace{2cm}
  \begin{tikzpicture}[baseline={(current bounding box.center)}]
    \useasboundingbox (-2,-2) rectangle (2,2);
    \begin{pgfonlayer}{edgelayer}
      \filldraw[fill=colD!50!white, draw=black, line width=1pt] (0,0) circle (0.8cm);
      \draw[ndi] (90:1.6) -- (90:0.8);
      \draw[ndi] (215:0.8) -- (215:1.6);
      \draw[ndi] (325:0.8) -- (325:1.6);
      \draw[diA,dashed] (70:1.4) arc (200:240:3);
      \draw[diA,dashed] (1,-1.2) arc (70:110:3);
      \draw (270:1.1) node[anchor=north] {$Q$};
      \draw (25:1.1) node[anchor=south west] {$P$};
    \end{pgfonlayer}
  \end{tikzpicture}
  \caption{External momenta asignment for auxiliary and symmetric point
    integrals(left) and for the large momentum expansion procedure(right).}
  \label{fig:kin3pt}
\end{figure}
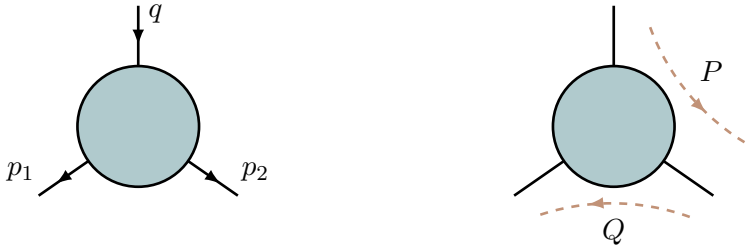

Before providing details of the three-loop calculation, we need to review
techniques used at two loops. The first two-loop calculation at the symmetric
point~\cite{Chetyrkin:2000fd} used a large momentum expansion procedure. This technique
does not rely on any symmetric point integrals knowledge and requires only
results for massless propagator integrals. In subsequent two-loop
calculations~\cite{Almeida:2010ns,Gracey:2011pf,Gracey:2011fb,Gracey:2011zn,Gracey:2011zg},
authors adopted \emph{Integration-By-Parts}(IBP)~\cite{Tkachov:1981wb,Chetyrkin:1981qh} reduction to the minimal set of two-loop master integrals
known for a long time, even in general kinematics not restricted to the
symmetric point~\cite{Davydychev:1992xr,Usyukina:1994iw,Birthwright:2004kk}.
This way, the problem of the master integrals calculation had not appeared
explicitly in those works.

On the other hand, a very restricted subset of the three-loop master integrals
needed for analytical calculations at the symmetric
point~\cite{Bednyakov:2020ugu,Bednyakov:2020cdf} is
known~\cite{Usyukina:1993ch}.

Our starting point for three-loop master integrals calculation is a
paper~\cite{Chavez:2012kn}, where linear reducibility of the two-loop
three-point integrals with off-shell legs is demonstrated after an appropriate
variable change. Linear reducibility of three-loop integrals with off-shell
kinematics was indicated in the E.Panzer Ph.D. thesis~\cite{Panzer:2015ida}. Since symmetric
point integrals are a particular case of considered integrals, we expect them
to be expressed in terms of generalized polylogarithms(GPL) as a direct consequence
of the linear reducibility. The same property will have auxiliary integrals with
$p_1^2 = p_2^2$, but $q^2$ arbitrary, and we will be able to reduce differential
equations(DE) system for such integrals to the so-called $\ep$-form~\cite{Henn:2013pwa}.

As the primary calculation method, we solve differential equations system for
auxiliary integrals. Due to performance issues with the most complicated
integrals, we use direct integration in terms of GPLs for partial checks only.
Also, the method of direct integration requires construction of a new set of master
integrals without sub-divergencies, which is possible to achieve by shifting
space-time dimension and increasing some propagators
powers~\cite{vonManteuffel:2014qoa}. Conversion between the set of finite
integrals and the original basis is an additional complication of the method.
However, an alternative based on the DE system solution is also highly
nontrivial. We need to consider more complicated integrals depending on at least
one dimensionless variable and provide boundary conditions(BC) sufficient to fix all
needed integration constants in the DE solution.

The paper organized as follows: we define auxiliary topologies containing all
integrals of our interest in section~\ref{sec:auxtopdef}, we describe the
solution of differential equations for auxiliary integrals in
section~\ref{sec:DEauxInts}, we describe our approach to fixing boundary
conditions with a detailed two-loop example in section~\ref{sec:fixBC},
construction of the uniform transcendentality weight basis of integrals can be found
in section~\ref{sec:utBasis}, and in section~\ref{sec:miRes} we provide the list
of calculated master integrals. Actual results for the integrals are in the
supplementary materials to the paper.

\section{Notations for integrals topologies}
\label{sec:auxtopdef}

\begin{table}[t]
  \centering
  \begin{tabular*}{0.7\textwidth}{l @{\extracolsep{\fill}} rr}
    & $A_2$   & $B_2$ \\ \hline
    $P_1$ & $k_1 + p_1$ & $k_1 + p_1$ \\
    $P_2$ & $k_2 + p_1$ & $k_2 + p_1$ \\
    $P_3$ & $k_1 - p_2$ & $k_1 - p_2$\\
    $P_4$ & $k_2 - p_2$ & $k_1 - k_2 - p_2$\\
    $P_5$ & $k_1 - k_2$ & $k_1 - k_2$\\
    $P_6$ & $k_1$ & $k_2$ \\
    $P_7$ & $k_2$ & $k_1$ 
  \end{tabular*}
  \caption{Two-loop auxiliary topologies}
  \label{tab:aux2l}
\end{table}
To reduce many integrals in the course of
calculations~\cite{Bednyakov:2020cdf,Bednyakov:2020ugu}, we apply
IBP reduction to the small set of master integrals
considered in the present paper. To perform the reduction efficiently, we use
the Laporta algorithm~\cite{Laporta:2001dd} implemented in the package
\texttt{Reduze2}~\cite{vonManteuffel:2012np}. We define auxiliary topologies
containing the complete set of propagators to express all the appearing scalar
products for the algorithm's work. To uniquely identify an integral inside
topology, we use a vector of integer numbers corresponding to propagators'
powers.

For a vertex integral at the symmetric point, we assign external momenta according to
the left part of the Fig.~\ref{fig:kin3pt} and set $p_1^2=p_2^2=q^2=-1$. This
definition makes all integrals real, and self-energy integrals are in one-to-one
correspondence with integrals from the \texttt{MINCER}
package~\cite{Gorishnii:1989gt,Larin:1991fz}. Later, this property will be
helpful for the boundary conditions fixing procedure described in
section~\ref{sec:fixBC}.

At the one-loop level, we have a single topology $A_1$ with three propagators
\begin{equation}
  \label{eq:topoA1def}
  I_{A_1}[n_1,n_2,n_{3}]  = e^{\ep \gamma_E}\int \frac{d^D k_1}{i \pi^{D/2}} \frac{1}{((k_1+p_1)^2)^{n_1} ((k_1-p_2)^2)^{n_2} (k_1^2)^{n_3}}. 
\end{equation}
At the two-loop level, we have two seven-propagator topologies, $A_2$ and $B_2$ 
\begin{equation}
  \label{eq:topos2ldef}
  I_{X_2}[n_1,\dots,n_{7}] = e^{2 \ep \gamma_E}\int \frac{d^D k_1}{i \pi^{D/2}} \frac{d^D k_2}{i \pi^{D/2}} \frac{1}{(P_{1}^2)^{n_1} \dots (P_{7}^2)^{n_{7}}},
\end{equation}
where $X=\{A,B\}$ and propagators' momenta are defined explicitly in table~\ref{tab:aux2l}.
\begin{table}[!h]
  \centering
  \begin{tabular*}{\textwidth}{l @{\extracolsep{\fill}} rrr}
    & $A_3$ & $B_3$ & $C_3$ \\ \hline
    $P_1$ & $k_1$ & $k_1$ & $k_1$ \\
    $P_2$ & $k_2$ & $k_2$ & $k_2$ \\
    $P_3$ & $k_3$ & $k_3$ & $k_3$ \\
    $P_4$ & $k_1 - k_2$ & $k_1 - k_2$ & $k_1 - k_2$ \\
    $P_5$ & $k_1 - k_3$ & $k_1 - k_3$ & $k_1 - k_3$ \\
    $P_6$ & $k_2 - k_3$ & $k_1 - k_2 - k_3$ & $k_2 - k_3$ \\
    $P_7$ & $k_1 - p_1$ & $k_1 - p_1$ & $k_1 - k_3 - p_2$ \\
    $P_8$ & $k_1 - p_1 - p_2$ & $k_1 - p_1 - p_2$ & $k_1 - p_1 - p_2$ \\
    $P_9$ & $k_2 - p_1$ & $k_2 - p_1$ & $k_2 - p_1$ \\
    $P_{10}$ & $k_2 - p_1 - p_2$ & $k_2 - p_1 - p_2$ & $k_1 - k_2 - p_2$ \\
    $P_{11}$ & $k_3 - p_1$ & $k_3 - p_1$ & $k_3 - p_1$ \\
    $P_{12}$ & $k_3 - p_1 - p_2$ & $k_3 - p_1 - p_2$ & $k_3 - p_1 - p_2$ 
  \end{tabular*}
  \caption{Three-loop auxiliary topologies}
  \label{tab:aux3l}
\end{table}
At three-loops, we have three topologies, $A_3$, $B_3$, and $C_3$, each with 12 propagators
\begin{equation}
  \label{eq:topos3ldef}
  I_{X_3}[n_1,\dots,n_{12}] = e^{3 \ep \gamma_E} \int \frac{d^D k_1}{i \pi^{D/2}}\frac{d^D k_2}{i \pi^{D/2}}\frac{d^D k_3}{i \pi^{D/2}}\frac{1}{(P_{1}^2)^{n_1} \dots (P_{12}^2)^{n_{12}}}.
\end{equation}
Table~\ref{tab:aux3l} provides explicit expressions for momenta $P_i$ for each of the three-loop topologies with $X=\{A,B,C\}$.
\begin{table}[!h]
  \centering
  \begin{tabular}{c||c|c||c|c|c}
    $A_1$ & $A_2$ & $B_2$ & $A_3$ & $B_3$ & $C_3$ \\ \hline
    2     & 7     &  1   & 38 & 12  & 1
  \end{tabular}
  \caption{Distribution of the number of integrals over topologies.}
  \label{tab:minum}
\end{table}

The number of unique master integrals after IBP reduction are given in
table~\ref{tab:minum}. We present all master integrals graphs in
tables~\ref{tab:oneloopints},~\ref{tab:twoloopints} and~\ref{tab:threeloopints}.

\section{Differential equations for auxiliary integrals}
\label{sec:DEauxInts}
As stated in the introduction, our main calculation method is a solution of the
DE system for specially constructed auxiliary integrals. We consider integrals
similar to original three-point ones, but now with $q^2$ not fixed, and introduce
scaleless variable $x$ with $q^2 = x\, p_1^2$. At first glance, this makes us
calculate more complex integrals than the initial set. However, the advantage is
that now we have access to the DE system, connecting point $x=1$ with any other
point, where boundary conditions can be constructed easily. In our calculation,
we consider the limit $x \to 0$ for fixing boundary conditions and the limit $x
\to \infty$ for checks.

We analyze the same topologies as before, but now with $q^2$ not fixed for
auxiliary integrals. After performing IBP reduction and identification of master
integrals for each of topologies, we have the following number of master
integrals: 3 for $A_1$, 12 for $A_2$, 11 for $B_2$, 88 for $A_3$ and 91 for
$B_3$. Since for topology $C_3$ at the symmetric point, we have only one integral,
which factorizes into lower-loop integrals from topologies $A_1$ and $B_2$, we
do not consider it here.

To differentiate master integrals in variable $x$ and reduce the result back to master integrals, we make use of the package
\texttt{Reduze 2}~\cite{vonManteuffel:2012np}. To convert the obtained DE system to the $\ep$-form with first apply a variable
change $x = -\frac{(z-1)^2}{z}$. In a new variable $z$, singular points of the
DE system correspond to the following set of singular points in an old variable
$x$:
\begin{equation}
  \label{eq:zLimsDef}
  z \to e^{\pm \frac{i \pi}{3}} : x \to 1,
  \quad
  z \to 1 : x \to 0,
  \quad
  z \to 0 : x \to \infty,
  \quad
  z \to -1 : x \to 4.
\end{equation}
The point $z = e^{\pm \frac{i \pi}{3}}$ is a solution we are interested in,
limit $z = 1$ we use in section~\ref{sec:fixBC} as boundary conditions to fix
integration constants in the DE solution. The solution in the point $z = 0$ is
used to derive known results for the there-loop massless form-factor type
integrals~\cite{Gehrmann:2010ue,Lee:2010cga,vonManteuffel:2015gxa}, which is necessary for
checking obtained results. We do not consider the point $z = -1$ in the present
paper.

After the variable change, the system of differential equations for the vector
of integrals $\vec{f}(z,\ep)$ for each of the considered topologies has the following
form
\begin{equation}
  \label{eq:deMatOrig}
  \partial_z  \vec{f}(z,\ep) = M(z,\ep) \vec{f}(z,\ep).
\end{equation}

With matrix $M$ having only a limited set of $\ep$-independent singularities of
the finite order in the set of points $S =\{0,1,-1,\lambda,\lambda^*\}$, where $\lambda = \sru$ and $\lambda^*$ is
its complex conjugate.
\begin{equation}
  \label{eq:MsingDE}
 M(z,\ep) =  \sum\limits_{z_j \in S} \sum\limits_{k \ge 0} \frac{M_{k}^{(z_j)}(\ep)}{(z-z_j)^{k+1}} + \sum\limits_{k \ge 0} z^k M_k(\ep)  
\end{equation}
This form of the DE system is ideally suited for further
conversion to the canonical form~\cite{Henn:2013pwa}. Only one complication is the appearance of
sixth-roots of unity residues at the three-loop order. For one and two-loop
integrals, the DE systems are singular only in $z=\{0,1,-1\}$.

Following the strategy from~\cite{Henn:2013pwa} and with
algorithm~\cite{Lee:2014ioa} implemented in the publically available package
\texttt{epsilon}~\cite{Prausa:2017ltv}, we convert the original DE system to
another DE system for a new basis of integrals $\vec{g}(z,\ep) = T^{-1}(z,\ep)
\vec{f}(z,\ep)$
\begin{equation}
  \label{eq:deMatEpForm}
  \partial_z \vec{g}(z,\ep) = \ep \left[
    \frac{A_{0}}{z}
    + \frac{A_{1}}{z-1}
    + \frac{A_{-1}}{z+1}
    + \frac{A_{\lambda}}{z-\lambda}
    + \frac{A_{\lambda^*}}{z-\lambda^*}  \right] \vec{g}(z,\ep).
\end{equation}
Obtained DE system matrix has only Fuchsian singularities, and $\ep$ dependence completely factorizes.
Applying \texttt{epsilon} for reduction of the original system to the
form~\eqref{eq:deMatEpForm}, we perform all the steps of the
algorithm~\cite{Lee:2014ioa} except the last one. Due to the performance issues,
we were forced to find a constant matrix transforming the system to the form
with $\ep$ factorized manually. Another reason to make the last step manually is
that the found transformation matrix is not unique, as not unique the DE system
matrix in $\ep$-form. This freedom in the transformation matrix and related
freedom in the $\ep$-form matrix we use in section \ref{sec:utBasis}, where we
construct a new basis of integrals with uniform transcendentality(UT) weight.

The differential equations system~\eqref{eq:deMatEpForm} has an excellent
property that, after expanding all the master integrals in $\ep$, the
differential equations for the series coefficients $\vec{g}_i(z)$ with
$\vec{g}(z,\ep) = \sum \vec{g}_i(z)\ep^i$ decouple entirely, and the solution for the
particular coefficient has the form:
\begin{equation}
  \label{eq:deSolEp}
  \vec{g}_{i}(z) = \int_{0}^{z} d y\left[     \frac{A_{0}}{y}
    + \frac{A_{1}}{y-1}
    + \frac{A_{-1}}{y+1}
    + \frac{A_{\lambda}}{y-\lambda}
    + \frac{A_{\lambda^*}}{y-\lambda^*}  \right] \vec{g}_{i-1}(y) + \vec{r}_{i}
\end{equation}

To perform integration, we utilize GPLs~\cite{Goncharov:2001iea} integration
rules, and starting with some order $n$ of $\ep$-expansion where all $\vec{g}_n
\equiv 0$ we proceed by induction.
At the next step we associate integration constants with zero weight GPLs, since
$G(;z) \equiv 1$. Singularities of the system~\eqref{eq:deMatEpForm} define the alphabet of GPLs. In
the examined case, we have a subset of the full six-root-of-unity alphabet
considered in~\cite{Henn:2015sem}, and for lower loop orders, the alphabet of HPLs
containing only  $\{0,1,-1\}$ is enough. GPLs are directly integrated with
\begin{equation}
  \label{eq:GPLint}
  G(a_1,\dots,a_n;z) = \int_{0}^{z} \frac{d t}{t-a_1} G(a_2,\dots,a_n;t).
\end{equation}

Each order of $\ep$-expansion of the constructed solution is built from the GPLs
of the uniform transcendental weight multiplied by some unknown constants. In
the general case, actual values of integration constants $\vec{r}_i$ can spoil
the uniform transcendentality(UT)
structure of the solution, and we address the question of restoring UT expansion
in section~\ref{sec:utBasis}. Our next step is to fix integration constants $\vec{r}_i$ by
providing appropriate boundary conditions.
\section{Fixing boundary conditions}
\label{sec:fixBC}

To fix integration constants in the solution of DE from the previous section, we
consider its behavior near the point $z=1$. Since the point $z=1$ is a singular
point of the DE system~\eqref{eq:deMatEpForm}, we can not take the limit
directly due to the logarithmic singularities $\log (1-z)$ appearing in the
solution. Thanks to the Fuchsian form of the DE system~\eqref{eq:deMatEpForm},
the leading order term of expansion in a small variable $\zb = 1-z$ can be
constructed directly from the matrix residue $A_1$~\cite{Dulat:2014mda}\footnote{Higher orders of
expansion in $\zb$ can be constructed recursively, as shown
in~\cite{Lee:2017qql}.}:
\begin{equation}
  \label{eq:limz1sol}
  \lim_{\zb\to 0} \vec{g}(\ep,z) = \zb^{\ep A_1} \vec{c}(\ep) + \mathcal{O}(\zb).
\end{equation}
This leading term of the solution is exact in $\ep$, and vector of unknown
$\ep$-dependent functions $\vec{c}(\ep)$ has the same size as the vector of
master integrals $\vec{g}(\ep,z)$. Knowledge of the vector of functions
$\vec{c}(\ep)$ up to sufficiently high order in $\ep$ is enough to fix all
needed integration constants. The series solution in the
form~\eqref{eq:limz1sol} allows us to separate contribution from different
subgraphs by considering terms with different non-integer powers of $\zb$. It is
handy to extract contribution from the hard subgraph, accessible from the naive
Taylor expansion of the integrand under the integral sign. However, in the
considered case, naive expansion is not enough to fix all required integration
constants, and we need to consider asymptotic expansion, including all relevant
subgraphs. To calculate the vector of functions $\vec{c}(\ep)$, we develop a
highly automated setup based on the package
\texttt{EXP}~\cite{Harlander:1997zb,Seidensticker:1999bb} and a version of the
\texttt{MINCER}~\cite{Gorishnii:1989gt,Larin:1991fz} package keeping all
intermediate $\ep$-dependent expressions in unexpanded form.

Momenta routing used for the asymptotic expansion coefficients calculation is
given in Fig.~\ref{fig:kin3pt}(right). With this routing, our kinematics
constraints are as follows $P^2 = x\, Q^2$ and $P\cdot Q =\frac{x}{2} Q^2$, and
results of the expansion in the limit $x \to 0$ we can obtain from the
\emph{large momentum procedure}(LMP) considering $Q$ large. For more details on
the LMP, see~\cite{Harlander:1999cs} and references therein. Expansion of the
arbitrary three-point vertex integral with external momenta $P,Q$ in large
momentum $Q$ has the form:
\begin{equation}
  \label{eq:expSeriesForm}
  J(P,Q) \sim \sum_{S} \left(\frac{P^2}{Q^2}\right)^{\ep S} \sum\limits_{i,j} a_{i,j}^S
  \left(\frac{P \cdot Q}{Q^2}\right)^i \left(\frac{P^2}{Q^2}\right)^j
\end{equation}
To fix coefficients $a_{i,j}^S$ of the expansion, we use setup based on the
package \texttt{EXP}. First, using \texttt{EXP}, we identify all needed
subgraphs to expand in large external momenta, then we expand to sufficiently
high order in the large momentum $Q$, further we perform tensor reduction to
separate powers of the scalar products $Q^2$ and $P\cdot Q$ from loop integrals
dependent on a single external momenta $Q$. The remaining two-point integrals
are calculated with \texttt{MINCER}, which is fast enough to provide us with
expansion terms to high order in variable $\zb$. We determine coefficient
$a_{i,j}$ as a linear combination of massless propagator-type master integrals,
these integrals we can expand later up to the required order in $\ep$.

After conversion of obtained expansions back with $T^{-1}$, we truncate the
series to get the leading term for $\vec{g}$ and fix $\vec{c}(\ep)$ from
Eq.~\eqref{eq:limz1sol}. Contributions of different subgraphs from different
integrals to the same set of functions $\vec{c}$ provide us with a strong check
on the validity of the procedure.

Now expanding both solution of the system \eqref{eq:deMatEpForm} in $z\to 1$ and
leading order solution \eqref{eq:limz1sol} with $\vec{c}(\ep)$ substituted, we can fix
all integration constants up to required order in $\ep$ simply expanding
\texttt{MINCER} master integrals deep enough in $\ep$.

As a real example, we consider the DE system for non-trivial two-loop integrals
\begin{equation}
  \label{eq:Jde2l}
  J_1 = I^{A2}_{0, 1, 0, 0, 1, 1, 0},
  \quad
  J_2 = I^{A2}_{0, 1, 1, 0, 1, 0, 0}, 
  \quad
  J_3 = I^{A2}_{0, 1, 1, 0, 1, 1, 0},
  \quad
  J_4 = I^{A2}_{0, 2, 1, 0, 1, 1, 0}.
\end{equation}
Integrals $\vec{J}$ are transformed to the canonical basis integrals $\vec{j} =
T^{-1}\vec{J}$ with the matrix
\begin{equation}
  \label{eq:mT}
  T =
  \begin{pmatrix}
    \frac{-2\ep^3}{(1 - 2\ep)(2 - 3\ep)(1 - 3\ep)} &  0 &  0&  0 \\
    0 & \frac{-2 \ep^3 (1 - z)^2}{( 1 - 2\ep)(2 - 3\ep)(1 - 3\ep)z} & 0 & 0\\
    0 &  0 & \frac{2\ep^2}{(1 - 2\ep)(1 - 3\ep)} & \frac{\ep^2z}{(1 - 2\ep)(1 - 3\ep)(1+ z)}\\
    0 & 0 & 0 & \frac{-\ep z}{(1 - z)(1 + z)}
  \end{pmatrix}.
\end{equation}
DE system for integrals $\vec{j}(z)$ now is in the $\ep$-form
\begin{equation}
  \label{eq:f2g}
  \partial_z \vec{j}(z) = \ep\left( \frac{A_0}{z} + \frac{A_1}{z-1} + \frac{A_{-1}}{z+1} \right) \vec{j}(z),
\end{equation}
with numeric matrices
\begin{equation}
  \label{eq:matsys}
A_0 = 
  \begin{pmatrix}
    0 & 0 & 0 & 0 \\
    0 & 2 & 0 & 0 \\
    0 & -1 & 1 & 0\\
    2 & 4 & -2 & -1
  \end{pmatrix}, \quad
  A_1 = 
  \begin{pmatrix}
    0 & 0 & 0 & 0 \\
    0 & -4 & 0 & 0 \\
    -1 & 0 & -1 & 0\\
    0 & 0 & 0 & -1
  \end{pmatrix}, \quad
A_{-1} = 
  \begin{pmatrix}
    0 & 0 & 0 & 0 \\
    0 & 0 & 0 & 0 \\
    0 & 0 & 0 & -\frac{1}{2}\\
    0 & 0 & 0 & 2
  \end{pmatrix}.
\end{equation}
Due to the Fuchsian form of the DE system, the leading order solution in the
limit $z \to 1$ is determined by the matrix exponent:
\begin{equation}
  \label{eq:liz1sol}
  \lim_{\zb\to 0} \vec{j}(\ep,z) = \zb^{\ep A_1} \vec{c}(\ep) + \mathcal{O}(\zb).
\end{equation}
With $\zb = 1-z$ it reads when we keep leading terms in $\zb$ only
\begin{equation}
  \label{eq:zb0Sol}
  j_1 = c_1,
  \quad
  j_2 = \zb^{-4\ep} c_2,
  \quad
  j_3 = -c_1 + \zb^{-\ep} \left(c_1 + c_3\right),
  \quad
  j_4 = \zb^{-\ep} c_4.
\end{equation}
Here all $c_i$ are functions of $\ep$, and we see that some of the functions
$c_i$ can be fixed independently from different integrals expansions stemming as
an additional cross check. In provided example $c_1$ can be fixed both from the
$j_1$ and $j_3$.
With \texttt{EXP} we obtain the following expansions for original integrals:
\begin{equation}
  \label{eq:Jexp2}
  J_1  = T_1, \quad
  J_2  = (-\zb^2)^{1-2\ep} T_1 + \mathcal{O}(\zb^3),  \quad
  J_3 =  \frac{2-3\ep}{\ep} T_1  + \mathcal{O}(\zb), \quad
  J_4  = \mathcal{O}(\zb)
\end{equation}
Where $T_1$ is the two-loop massless sunset integral with $Q^2 = -1$.
Above we provide only orders needed to fix all $c_i(\ep)$ entering \eqref{eq:zb0Sol}.
After multiplication with $T^{-1}$ we find the required functions:
\begin{equation}
  \label{eq:cFixSol}
  c_1 = - \frac{(1-2\ep)(2-3\ep)(1-3\ep)}{2 \ep^3} T_1,
  \quad
  c_2 = (-1)^{1-2\ep}c_1,
  \quad
  c_3 = - c_1,
  \quad
  c_4 = 0
\end{equation}


After fixing the required integration constants, the obtained solution can be
checked by considering the limit $z \to 0$, corresponding to infinitely large $(q^2
\to \infty)$
external momentum squared. The hard subgraph's contribution in this limit allows
us to extract results for the massless form-factor integrals and compare them
with expressions known up to the three-loop order~\cite{Lee:2010cga,vonManteuffel:2015gxa}.

As before due to the Fuchsian form of the DE system, we can follow procedure
described in~\cite{Lee:2017qql} and construct general form of the expansion
in $z \to 0$

\begin{equation}
  \label{eq:solSmallZexp}
  g(z,\ep) = \sum\limits_{i}z^{a_i \ep}\sum\limits_{j=0} c_j(\ep) z^j
\end{equation}
Expanding integrals $\vec{g}(z)$ in $\ep$, we can find several first terms of a
Laurent expansion of $c_j(\ep)$ around $\ep=0$. With the help of the matrix $T$,
we can construct expansion for the original basis of integrals $\vec{f}$, which
we utilize to extract terms with specific noninteger powers of $z$. We can
extract the value $I_{\rm FF}$ for the massless form-factor integral with
$L$-loops and $E$ internal edges from the hard subgraph contribution of the
form:
\begin{equation}
  \label{eq:ffLim}
  J_{\rm hard} = (-z)^{E-2L +L \ep}(I_{\rm FF} + \mathcal{O}(z))
\end{equation}
From the present paper results, we found agreement with the results of the paper
\cite{vonManteuffel:2015gxa} for all master integrals up to the transcendental
weight six.

\tikzstyle{none}=[inner sep=0mm]
\tikzstyle{dot}=[fill=black, draw=black, shape=circle]
\tikzstyle{prop}=[-, draw=colA, line width=4pt]
\tikzstyle{legs}=[-, draw=colD, line width=4pt]

\begin{table}
  \centering
  \begin{tabular}{cc}
    
  \resizebox{2cm}{!}{\input figs/1-loop/A1_110.tikz}

    & 
  \resizebox{2cm}{!}{\input figs/1-loop/A1_111.tikz}
\\
    \icap{A1}{110}
    & \icap{A1}{111}
  \end{tabular}  
  \caption{One-loop master integrals}
  \label{tab:oneloopints}
\end{table}

\begin{table}
  \centering
  \begin{tabular}{cccc}
    
  \resizebox{2cm}{!}{\input figs/2-loop/A2_0110100.tikz}

    & 
  \resizebox{2cm}{!}{\input figs/2-loop/A2_0210110.tikz}

    & 
  \resizebox{2cm}{!}{\input figs/2-loop/A2_1101110.tikz}

    & 
  \resizebox{2cm}{!}{\input figs/2-loop/A2_1111010.tikz}
\\
    \icap{A2}{0110100}
    & \icap{A2}{0210110}
    & \icap{A2}{1101110}
    & \icap{A2}{1111010}\\

  \resizebox{2cm}{!}{\input figs/2-loop/A2_0110110.tikz}

    & 
  \resizebox{2cm}{!}{\input figs/2-loop/A2_1101010.tikz}

    & 
  \resizebox{2cm}{!}{\input figs/2-loop/A2_1111000.tikz}

    & 
  \resizebox{2cm}{!}{\input figs/2-loop/B2_1111110.tikz}
\\
    \icap{A2}{0110110}
    & \icap{A2}{1101010}
    & \icap{A2}{1111000}
    & \icap{B2}{1111110}
  \end{tabular}
  \caption{Two-loop master integrals}
  \label{tab:twoloopints}
\end{table}

\section{Basis of integrals with uniform transcendentality weight}
\label{sec:utBasis}

The main problem with the already obtained solution is that its expansion
coefficients lack the UT property. For practical applications like~\cite{Bednyakov:2020cdf,Bednyakov:2020ugu},
we need to expand some of the integrals to high order in $\ep$ corresponding to the
transcendental weight seven. Only part of the integrals with GPLs up to the
transcendental weight six enters the final result, and all GPLs with
transcendental weight seven cancel in the sum. At intermediate steps,
manipulations with a weight seven GPLs are highly complicated, since reduction
rules are available up to the weight six only~\cite{Henn:2015sem}.

Possible solution to the problem is a new basis of UT weight functions,
which need to be known up to the transcendentality weight six to express results
containing GPLs up to the weight six. Explicitly we want to find a basis of pure
functions $\vec{U} = T_{\rm UT}^{-1} \vec{f}$, with $\ep$-expansion in the form $\vec{U} =
\sum_{j=0}^{6} \vec{u}_{j} \ep^j + \mathcal{O}(\ep^7)$, where each coefficient $\vec{u}_j$ has
uniform transcendental weight $j$.

The UT integral basis construction is a complicated task even with special tools
designed to attack the problem~\cite{Dlapa:2020cwj,Henn:2020lye}. Fortunately,
with already available results we can easily avoid all these difficulties.

First, we notice that if the integrals with arbitrary $z$ have UT weight, this
property also holds for the integrals at the symmetric point($z \to \sru$). The
weight of GPLs and all $z$-dependent prefactors do not change transcendentality
after taking the limit, and we can focus on constructing UT basis for auxiliary
integrals.

Second, due to the UT property, the DE system for UT integrals is also in
$\ep$-form. Our goal is to find a transformation matrix between canonical master
integrals from section~\ref{sec:DEauxInts} and the new UT basis. As mentioned in
section~\ref{sec:DEauxInts}, the last step of reduction to the $\ep$-form has
afreedom, and we are looking for the transformation to UT basis by varying
these parameters.

Another important observation is that for the $z$-dependent integrals with UT
weight, this property also holds when considering their expansion near singular
points. For example leading terms of expansion in $z \to 1$ used for fixing
boundary conditions in section \ref{sec:fixBC}.
\begin{equation}
  \label{eq:utExp}
  J_{\rm UT}(\zb) \sim \sum\limits_{i}\zb^{a_i \ep}C_i
\end{equation}
Since $\ep$-expansion of the prefactor $\zb^{a_i \ep}$ in \eqref{eq:utExp} has transcendental weight zero, all coefficients $C_i$ for
expansion of the UT integral $J_{\rm UT}$ also have UT weight. In considered limit, all
coefficients $C_i$ are built from massless propagator integrals only. After
reduction to master integrals, most of them are known for arbitrary $D$, and for
the remaining, transformation to the UT basis is known from~\cite{Lee:2011jt}. In this way,
we find transformation matrix $T_{\rm UT}$ by analyzing expansion coefficients in the limit
$z \to 1$.

All UT integrals constructed in this way are regular in
the limit $z \to \sru$ and going back to the original integrals and then
reducing to the basis of symmetric point integrals, we obtain representation for
integrals of our interest in terms of pure functions.

\begin{table}
  \centering
  \begin{tabular}{cccccc}
    
  \resizebox{2cm}{!}{\input figs/3-loop/A3_001010111100.tikz}

    & 
  \resizebox{2cm}{!}{\input figs/3-loop/A3_001101100000.tikz}

    & 
  \resizebox{2cm}{!}{\input figs/3-loop/A3_001101110000.tikz}

    & 
  \resizebox{2cm}{!}{\input figs/3-loop/A3_001110011000.tikz}

    & 
  \resizebox{2cm}{!}{\input figs/3-loop/A3_001110021000.tikz}

    & 
  \resizebox{2cm}{!}{\input figs/3-loop/A3_001111011000.tikz}

    \\
    \icap{A3}{001010111100}
    & \icap{A3}{001101100000}
    & \icap{A3}{001101110000}
    & \icap{A3}{001110011000}
    & \icap{A3}{001110021000}
    & \icap{A3}{001111011000}
    \\
    
  \resizebox{2cm}{!}{\input figs/3-loop/A3_002010111100.tikz}

    & 
  \resizebox{2cm}{!}{\input figs/3-loop/A3_002101110000.tikz}

    & 
  \resizebox{2cm}{!}{\input figs/3-loop/A3_002111011000.tikz}

    & 
  \resizebox{2cm}{!}{\input figs/3-loop/A3_011000110110.tikz}

    & 
  \resizebox{2cm}{!}{\input figs/3-loop/A3_011010011000.tikz}

    & 
  \resizebox{2cm}{!}{\input figs/3-loop/A3_011010101000.tikz}

    \\
    \icap{A3}{002010111100}
    & \icap{A3}{002101110000}
    & \icap{A3}{002111011000}
    & \icap{A3}{011000110110}
    & \icap{A3}{011010011000}
    & \icap{A3}{011010101000}
    \\
    
  \resizebox{2cm}{!}{\input figs/3-loop/A3_011010111000.tikz}

    & 
  \resizebox{2cm}{!}{\input figs/3-loop/A3_011011111000.tikz}

    & 
  \resizebox{2cm}{!}{\input figs/3-loop/A3_011011211000.tikz}

    & 
  \resizebox{2cm}{!}{\input figs/3-loop/A3_011100110110.tikz}

    & 
  \resizebox{2cm}{!}{\input figs/3-loop/A3_011110011000.tikz}

    & 
  \resizebox{2cm}{!}{\input figs/3-loop/A3_011110100000.tikz}

    \\
    \icap{A3}{011010111000}
    & \icap{A3}{011011111000}
    & \icap{A3}{011011211000}
    & \icap{A3}{011100110110}
    & \icap{A3}{011110011000}
    & \icap{A3}{011110100000}
    \\
    
  \resizebox{2cm}{!}{\input figs/3-loop/A3_011110110000.tikz}

    & 
  \resizebox{2cm}{!}{\input figs/3-loop/A3_011110111000.tikz}

    & 
  \resizebox{2cm}{!}{\input figs/3-loop/A3_011111110110.tikz}

    & 
  \resizebox{2cm}{!}{\input figs/3-loop/A3_012010111000.tikz}

    & 
  \resizebox{2cm}{!}{\input figs/3-loop/A3_012110011000.tikz}

    & 
  \resizebox{2cm}{!}{\input figs/3-loop/A3_021011111000.tikz}

    \\
    \icap{A3}{011110110000}
    & \icap{A3}{011110111000}
    & \icap{A3}{011111110110}
    & \icap{A3}{012010111000}
    & \icap{A3}{012110011000}
    & \icap{A3}{021011111000}
    \\
    
  \resizebox{2cm}{!}{\input figs/3-loop/A3_021110011000.tikz}

    & 
  \resizebox{2cm}{!}{\input figs/3-loop/A3_021110110000.tikz}

    & 
  \resizebox{2cm}{!}{\input figs/3-loop/A3_021111110110.tikz}

    & 
  \resizebox{2cm}{!}{\input figs/3-loop/A3_101001111000.tikz}

    & 
  \resizebox{2cm}{!}{\input figs/3-loop/A3_101011011000.tikz}

    & 
  \resizebox{2cm}{!}{\input figs/3-loop/A3_110011011000.tikz}

    \\
    \icap{A3}{021110011000}
    & \icap{A3}{021110110000}
    & \icap{A3}{021111110110}
    & \icap{A3}{101001111000}
    & \icap{A3}{101011011000}
    & \icap{A3}{110011011000}
    \\
    
  \resizebox{2cm}{!}{\input figs/3-loop/A3_110011101000.tikz}

    & 
  \resizebox{2cm}{!}{\input figs/3-loop/A3_110011111000.tikz}

    & 
  \resizebox{2cm}{!}{\input figs/3-loop/A3_111000011010.tikz}

    & 
  \resizebox{2cm}{!}{\input figs/3-loop/A3_111000101010.tikz}

    & 
  \resizebox{2cm}{!}{\input figs/3-loop/A3_111000110110.tikz}

    & 
  \resizebox{2cm}{!}{\input figs/3-loop/A3_111000111010.tikz}

    \\
    \icap{A3}{110011101000}
    & \icap{A3}{110011111000}
    & \icap{A3}{111000011010}
    & \icap{A3}{111000101010}
    & \icap{A3}{111000110110}
    & \icap{A3}{111000111010}
    \\
    
  \resizebox{2cm}{!}{\input figs/3-loop/A3_111100011010.tikz}

    & 
  \resizebox{2cm}{!}{\input figs/3-loop/A3_210011011000.tikz}

    & 
  \resizebox{2cm}{!}{\input figs/3-loop/B3_001111100110.tikz}

    & 
  \resizebox{2cm}{!}{\input figs/3-loop/B3_010101110110.tikz}

    & 
  \resizebox{2cm}{!}{\input figs/3-loop/B3_011001110110.tikz}

    & 
  \resizebox{2cm}{!}{\input figs/3-loop/B3_011002110110.tikz}

    \\
    \icap{A3}{111100011010}
    & \icap{A3}{210011011000}
    & \icap{B3}{001111100110}
    & \icap{B3}{010101110110}
    & \icap{B3}{011001110110}
    & \icap{B3}{011002110110}
    \\
    
  \resizebox{2cm}{!}{\input figs/3-loop/B3_011111100110.tikz}

    & 
  \resizebox{2cm}{!}{\input figs/3-loop/B3_011111101010.tikz}

    & 
  \resizebox{2cm}{!}{\input figs/3-loop/B3_011111101110.tikz}

    & 
  \resizebox{2cm}{!}{\input figs/3-loop/B3_011111110110.tikz}

    & 
  \resizebox{2cm}{!}{\input figs/3-loop/B3_011211110110.tikz}

    & 
  \resizebox{2cm}{!}{\input figs/3-loop/B3_020101110110.tikz}

    \\
    \icap{B3}{011111100110}
    & \icap{B3}{011111101010}
    & \icap{B3}{011111101110}
    & \icap{B3}{011111110110}
    & \icap{B3}{011211110110}
    & \icap{B3}{020101110110}
    \\
    
  \resizebox{2cm}{!}{\input figs/3-loop/B3_021111100110.tikz}

    & 
  \resizebox{2cm}{!}{\input figs/3-loop/B3_021111110110.tikz}

    & 
  \resizebox{2cm}{!}{\input figs/3-loop/C3_110110111100.tikz}

    &
    &
    \\
      \icap{B3}{021111100110}
    & \icap{B3}{021111110110}
    & \icap{C3}{110110111100}
    &
    &
  \end{tabular}
  \caption{Three-loop master integrals}
  \label{tab:threeloopints}
\end{table}

\section{Results and conclusion}
\label{sec:miRes}

In addition to the described method, we use IBP reduction to the new basis of
finite integrals to follow the strategy described in the
paper~\cite{vonManteuffel:2014qoa}. With package
\texttt{HyperInt}~\cite{Panzer:2014caa}, we find an analytical solution for
several integrals, except the most complicated, which is in complete agreement
with results of the previuos sections. All
integrals calculated in the paper were checked numerically with the sector
decomposition approach implemented in the package
\texttt{pySecDec}~\cite{Borowka:2017idc}.

In tables~\ref{tab:oneloopints}, \ref{tab:twoloopints} and \ref{tab:threeloopints},
we present all master integrals needed for calculations at the symmetric point
up to the three-loop order. Analytical results for integrals can be found in
supplementary files to the paper and contain a transformation matrix to the
basis of UT weight functions and expansion of latter up to the
transcendental weight six.

Up to the three-loop level, expansion of the basis functions is expressed in
terms of HPLs with the sixth-root of unity argument. Instead of the basis
constructed in paper~\cite{Kniehl:2017ikj}, which has an attractive feature,
that reduction rules for real(imaginary) part of HPLs contain only real(imaginary) parts of basis
functions. We construct a new
basis allowing us to present results in a more compact form. Also, our basis
contains as basis elements products of lower weight functions, which simplifies
products of lower-loop integrals during renormalization. The total number of
elements $N_w$ at weight $w$ is the same $N_w = 2 F_{2w}$, where $F_n$ is the
$n$-th Fibonacci number and explicitly is given by $N_w=\{2, 6, 16, 42, 110,
288\}$ for the weights considered in the paper.

In the paper, we calculated all massless three-point master integrals at the
symmetric point. As the solution method, we apply differential equations for
specially constructed auxiliary integrals and fix boundary conditions from the
large momentum expansion. Results for integrals are expressed through the basis
of functions with uniform transcendental weight. We provide expansion in $\ep$
for these functions in terms of the HPLs with the sixth root of unity argument
up to the transcendental weight six. The obtained result can find its
application in future calculations in the RI/SMOM scheme and as the boundary
conditions for three-point integrals in more general kinematics.

\acknowledgments

We gratefully acknowledge Alexander Bednyakov for collaboration
on~\cite{Bednyakov:2020cdf,Bednyakov:2020ugu}. We thank Roman Lee for his
valuable suggestions and discussions. We are grateful to the Joint
Institute for Nuclear Research for letting us use their
supercomputer``Govorun''. This work was supported by Russian Science
Foundation, grant 20-12-00205.

\bibliography{isym3pt3l}

\providecommand{\href}[2]{#2}\begingroup\raggedright\begin{thebibliography}{10}

\bibitem{Baikov:2016tgj}
P.A.~Baikov, K.G.~Chetyrkin and J.H.~Kühn, \emph{{Five-Loop Running of the QCD
  coupling constant}},
  \href{https://doi.org/10.1103/PhysRevLett.118.082002}{\emph{Phys. Rev. Lett.}
  {\bfseries 118} (2017) 082002}
  [\href{https://arxiv.org/abs/1606.08659}{{\ttfamily 1606.08659}}].

\bibitem{Luthe:2017ttg}
T.~Luthe, A.~Maier, P.~Marquard and Y.~Schroder, \emph{{The five-loop Beta
  function for a general gauge group and anomalous dimensions beyond Feynman
  gauge}}, \href{https://doi.org/10.1007/JHEP10(2017)166}{\emph{JHEP}
  {\bfseries 10} (2017) 166}
  [\href{https://arxiv.org/abs/1709.07718}{{\ttfamily 1709.07718}}].

\bibitem{Herzog:2017ohr}
F.~Herzog, B.~Ruijl, T.~Ueda, J.A.M.~Vermaseren and A.~Vogt, \emph{{The
  five-loop beta function of Yang-Mills theory with fermions}},
  \href{https://doi.org/10.1007/JHEP02(2017)090}{\emph{JHEP} {\bfseries 02}
  (2017) 090} [\href{https://arxiv.org/abs/1701.01404}{{\ttfamily
  1701.01404}}].

\bibitem{Chetyrkin:2017bjc}
K.G.~Chetyrkin, G.~Falcioni, F.~Herzog and J.A.M.~Vermaseren, \emph{{Five-loop
  renormalisation of QCD in covariant gauges}},
  \href{https://doi.org/10.1007/JHEP12(2017)006, 10.3204/PUBDB-2018-02123,
  10.1007/JHEP10(2017)179}{\emph{JHEP} {\bfseries 10} (2017) 179}
  [\href{https://arxiv.org/abs/1709.08541}{{\ttfamily 1709.08541}}].

\bibitem{tHooft:1973mfk}
G.~'t~Hooft, \emph{{Dimensional regularization and the renormalization group}},
  \href{https://doi.org/10.1016/0550-3213(73)90376-3}{\emph{Nucl. Phys.}
  {\bfseries B61} (1973) 455}.

\bibitem{Celmaster:1979km}
W.~Celmaster and R.J.~Gonsalves, \emph{{The Renormalization Prescription
  Dependence of the QCD Coupling Constant}},
  \href{https://doi.org/10.1103/PhysRevD.20.1420}{\emph{Phys. Rev.} {\bfseries
  D20} (1979) 1420}.

\bibitem{Gracey:2011pf}
J.A.~Gracey, \emph{{Three loop QCD MOM beta-functions}},
  \href{https://doi.org/10.1016/j.physletb.2011.04.052}{\emph{Phys. Lett.}
  {\bfseries B700} (2011) 79}
  [\href{https://arxiv.org/abs/1104.5382}{{\ttfamily 1104.5382}}].

\bibitem{Almeida:2010ns}
L.G.~Almeida and C.~Sturm, \emph{{Two-loop matching factors for light quark
  masses and three-loop mass anomalous dimensions in the RI/SMOM schemes}},
  \href{https://doi.org/10.1103/PhysRevD.82.054017}{\emph{Phys. Rev.}
  {\bfseries D82} (2010) 054017}
  [\href{https://arxiv.org/abs/1004.4613}{{\ttfamily 1004.4613}}].

\bibitem{Gracey:2011fb}
J.A.~Gracey, \emph{{RI'/SMOM scheme amplitudes for quark currents at two
  loops}}, \href{https://doi.org/10.1140/epjc/s10052-011-1567-8}{\emph{Eur.
  Phys. J.} {\bfseries C71} (2011) 1567}
  [\href{https://arxiv.org/abs/1101.5266}{{\ttfamily 1101.5266}}].

\bibitem{Gracey:2011zn}
J.A.~Gracey, \emph{{Two loop renormalization of the n = 2 Wilson operator in
  the RI'/SMOM scheme}},
  \href{https://doi.org/10.1007/JHEP03(2011)109}{\emph{JHEP} {\bfseries 03}
  (2011) 109} [\href{https://arxiv.org/abs/1103.2055}{{\ttfamily 1103.2055}}].

\bibitem{Gracey:2011zg}
J.A.~Gracey, \emph{{Amplitudes for the n = 3 moment of the Wilson operator at
  two loops in the RI/'SMOM scheme}},
  \href{https://doi.org/10.1103/PhysRevD.84.016002}{\emph{Phys. Rev.}
  {\bfseries D84} (2011) 016002}
  [\href{https://arxiv.org/abs/1105.2138}{{\ttfamily 1105.2138}}].

\bibitem{Bednyakov:2020ugu}
A.~Bednyakov and A.~Pikelner, \emph{{Quark masses: N3LO bridge from ${\rm
  RI/SMOM}$ to ${\rm \overline{MS}}$ scheme}},
  \href{https://doi.org/10.1103/PhysRevD.101.091501}{\emph{Phys. Rev.}
  {\bfseries D101} (2020) 091501}
  [\href{https://arxiv.org/abs/2002.12758}{{\ttfamily 2002.12758}}].

\bibitem{Bednyakov:2020cdf}
A.~Bednyakov and A.~Pikelner, \emph{{Four-loop QCD MOM beta functions from the
  three-loop vertices at the symmetric point}},
  \href{https://doi.org/10.1103/PhysRevD.101.071502}{\emph{Phys. Rev.}
  {\bfseries D101} (2020) 071502}
  [\href{https://arxiv.org/abs/2002.02875}{{\ttfamily 2002.02875}}].

\bibitem{Kniehl:2020nhw}
B.A.~Kniehl and O.L.~Veretin, \emph{{Moments $n=2$ and $n=3$ of the Wilson
  twist-two operators at three loops in the RI${}'$/SMOM scheme}},
  \href{https://doi.org/10.1016/j.nuclphysb.2020.115229}{\emph{Nucl. Phys.}
  {\bfseries B961} (2020) 115229}
  [\href{https://arxiv.org/abs/2009.11325}{{\ttfamily 2009.11325}}].

\bibitem{Kniehl:2020sgo}
B.A.~Kniehl and O.L.~Veretin, \emph{{Bilinear quark operators in the RI/SMOM
  scheme at three loops}},
  \href{https://doi.org/10.1016/j.physletb.2020.135398}{\emph{Phys. Lett.}
  {\bfseries B804} (2020) 135398}
  [\href{https://arxiv.org/abs/2002.10894}{{\ttfamily 2002.10894}}].

\bibitem{Chetyrkin:2000fd}
K.G.~Chetyrkin and T.~Seidensticker, \emph{{Two loop QCD vertices and three
  loop MOM beta functions}},
  \href{https://doi.org/10.1016/S0370-2693(00)01217-X}{\emph{Phys. Lett.}
  {\bfseries B495} (2000) 74}
  [\href{https://arxiv.org/abs/hep-ph/0008094}{{\ttfamily hep-ph/0008094}}].

\bibitem{Tkachov:1981wb}
F.V.~Tkachov, \emph{{A Theorem on Analytical Calculability of Four Loop
  Renormalization Group Functions}},
  \href{https://doi.org/10.1016/0370-2693(81)90288-4}{\emph{Phys. Lett.}
  {\bfseries 100B} (1981) 65}.

\bibitem{Chetyrkin:1981qh}
K.G.~Chetyrkin and F.V.~Tkachov, \emph{{Integration by Parts: The Algorithm to
  Calculate beta Functions in 4 Loops}},
  \href{https://doi.org/10.1016/0550-3213(81)90199-1}{\emph{Nucl. Phys.}
  {\bfseries B192} (1981) 159}.

\bibitem{Davydychev:1992xr}
A.I.~Davydychev, \emph{{Recursive algorithm of evaluating vertex type Feynman
  integrals}}, {\emph{J. Phys.} {\bfseries A25} (1992) 5587}.

\bibitem{Usyukina:1994iw}
N.I.~Usyukina and A.I.~Davydychev, \emph{{New results for two loop off-shell
  three point diagrams}},
  \href{https://doi.org/10.1016/0370-2693(94)90874-5}{\emph{Phys. Lett.}
  {\bfseries B332} (1994) 159}
  [\href{https://arxiv.org/abs/hep-ph/9402223}{{\ttfamily hep-ph/9402223}}].

\bibitem{Birthwright:2004kk}
T.G.~Birthwright, E.W.N.~Glover and P.~Marquard, \emph{{Master integrals for
  massless two-loop vertex diagrams with three offshell legs}},
  \href{https://doi.org/10.1088/1126-6708/2004/09/042}{\emph{JHEP} {\bfseries
  09} (2004) 042} [\href{https://arxiv.org/abs/hep-ph/0407343}{{\ttfamily
  hep-ph/0407343}}].

\bibitem{Usyukina:1993ch}
N.I.~Usyukina and A.I.~Davydychev, \emph{{Exact results for three and four
  point ladder diagrams with an arbitrary number of rungs}},
  \href{https://doi.org/10.1016/0370-2693(93)91118-7}{\emph{Phys. Lett.}
  {\bfseries B305} (1993) 136}.

\bibitem{Chavez:2012kn}
F.~Chavez and C.~Duhr, \emph{{Three-mass triangle integrals and single-valued
  polylogarithms}}, \href{https://doi.org/10.1007/JHEP11(2012)114}{\emph{JHEP}
  {\bfseries 11} (2012) 114} [\href{https://arxiv.org/abs/1209.2722}{{\ttfamily
  1209.2722}}].

\bibitem{Panzer:2015ida}
E.~Panzer, \emph{{Feynman integrals and hyperlogarithms}}, Ph.D. thesis,
  Humboldt U., 2015.
\newblock \href{https://arxiv.org/abs/1506.07243}{{\ttfamily 1506.07243}}.
\newblock 10.18452/17157.

\bibitem{Henn:2013pwa}
J.M.~Henn, \emph{{Multiloop integrals in dimensional regularization made
  simple}}, \href{https://doi.org/10.1103/PhysRevLett.110.251601}{\emph{Phys.
  Rev. Lett.} {\bfseries 110} (2013) 251601}
  [\href{https://arxiv.org/abs/1304.1806}{{\ttfamily 1304.1806}}].

\bibitem{vonManteuffel:2014qoa}
A.~von Manteuffel, E.~Panzer and R.M.~Schabinger, \emph{{A quasi-finite basis
  for multi-loop Feynman integrals}},
  \href{https://doi.org/10.1007/JHEP02(2015)120}{\emph{JHEP} {\bfseries 02}
  (2015) 120} [\href{https://arxiv.org/abs/1411.7392}{{\ttfamily 1411.7392}}].

\bibitem{Laporta:2001dd}
S.~Laporta, \emph{{High precision calculation of multiloop Feynman integrals by
  difference equations}}, \href{https://doi.org/10.1016/S0217-751X(00)00215-7,
  10.1142/S0217751X00002157}{\emph{Int. J. Mod. Phys.} {\bfseries A15} (2000)
  5087} [\href{https://arxiv.org/abs/hep-ph/0102033}{{\ttfamily
  hep-ph/0102033}}].

\bibitem{vonManteuffel:2012np}
A.~von Manteuffel and C.~Studerus, \emph{{Reduze 2 - Distributed Feynman
  Integral Reduction}},  \href{https://arxiv.org/abs/1201.4330}{{\ttfamily
  1201.4330}}.

\bibitem{Gorishnii:1989gt}
S.G.~Gorishnii, S.A.~Larin, L.R.~Surguladze and F.V.~Tkachov, \emph{{Mincer:
  Program for Multiloop Calculations in Quantum Field Theory for the
  Schoonschip System}},
  \href{https://doi.org/10.1016/0010-4655(89)90134-3}{\emph{Comput. Phys.
  Commun.} {\bfseries 55} (1989) 381}.

\bibitem{Larin:1991fz}
S.A.~Larin, F.V.~Tkachov and J.A.M.~Vermaseren, \emph{{The FORM version of
  MINCER}}, .

\bibitem{Gehrmann:2010ue}
T.~Gehrmann, E.W.N.~Glover, T.~Huber, N.~Ikizlerli and C.~Studerus,
  \emph{{Calculation of the quark and gluon form factors to three loops in
  QCD}}, \href{https://doi.org/10.1007/JHEP06(2010)094}{\emph{JHEP} {\bfseries
  06} (2010) 094} [\href{https://arxiv.org/abs/1004.3653}{{\ttfamily
  1004.3653}}].

\bibitem{Lee:2010cga}
R.N.~Lee, A.V.~Smirnov and V.A.~Smirnov, \emph{{Analytic Results for Massless
  Three-Loop Form Factors}},
  \href{https://doi.org/10.1007/JHEP04(2010)020}{\emph{JHEP} {\bfseries 04}
  (2010) 020} [\href{https://arxiv.org/abs/1001.2887}{{\ttfamily 1001.2887}}].

\bibitem{vonManteuffel:2015gxa}
A.~von Manteuffel, E.~Panzer and R.M.~Schabinger, \emph{{On the Computation of
  Form Factors in Massless QCD with Finite Master Integrals}},
  \href{https://doi.org/10.1103/PhysRevD.93.125014}{\emph{Phys. Rev.}
  {\bfseries D93} (2016) 125014}
  [\href{https://arxiv.org/abs/1510.06758}{{\ttfamily 1510.06758}}].

\bibitem{Lee:2014ioa}
R.N.~Lee, \emph{{Reducing differential equations for multiloop master
  integrals}}, \href{https://doi.org/10.1007/JHEP04(2015)108}{\emph{JHEP}
  {\bfseries 04} (2015) 108} [\href{https://arxiv.org/abs/1411.0911}{{\ttfamily
  1411.0911}}].

\bibitem{Prausa:2017ltv}
M.~Prausa, \emph{{epsilon: A tool to find a canonical basis of master
  integrals}}, \href{https://doi.org/10.1016/j.cpc.2017.05.026}{\emph{Comput.
  Phys. Commun.} {\bfseries 219} (2017) 361}
  [\href{https://arxiv.org/abs/1701.00725}{{\ttfamily 1701.00725}}].

\bibitem{Goncharov:2001iea}
A.B.~Goncharov, \emph{{Multiple polylogarithms and mixed Tate motives}},
  \href{https://arxiv.org/abs/math/0103059}{{\ttfamily math/0103059}}.

\bibitem{Henn:2015sem}
J.M.~Henn, A.V.~Smirnov and V.A.~Smirnov, \emph{{Evaluating Multiple
  Polylogarithm Values at Sixth Roots of Unity up to Weight Six}},
  \href{https://doi.org/10.1016/j.nuclphysb.2017.03.026}{\emph{Nucl. Phys.}
  {\bfseries B919} (2017) 315}
  [\href{https://arxiv.org/abs/1512.08389}{{\ttfamily 1512.08389}}].

\bibitem{Dulat:2014mda}
F.~Dulat and B.~Mistlberger, \emph{{Real-Virtual-Virtual contributions to the
  inclusive Higgs cross section at N3LO}},
  \href{https://arxiv.org/abs/1411.3586}{{\ttfamily 1411.3586}}.

\bibitem{Lee:2017qql}
R.N.~Lee, A.V.~Smirnov and V.A.~Smirnov, \emph{{Solving differential equations
  for Feynman integrals by expansions near singular points}},
  \href{https://doi.org/10.1007/JHEP03(2018)008}{\emph{JHEP} {\bfseries 03}
  (2018) 008} [\href{https://arxiv.org/abs/1709.07525}{{\ttfamily
  1709.07525}}].

\bibitem{Harlander:1997zb}
R.~Harlander, T.~Seidensticker and M.~Steinhauser, \emph{{Complete corrections
  of Order alpha alpha-s to the decay of the Z boson into bottom quarks}},
  \href{https://doi.org/10.1016/S0370-2693(98)00220-2}{\emph{Phys. Lett.}
  {\bfseries B426} (1998) 125}
  [\href{https://arxiv.org/abs/hep-ph/9712228}{{\ttfamily hep-ph/9712228}}].

\bibitem{Seidensticker:1999bb}
T.~Seidensticker, \emph{{Automatic application of successive asymptotic
  expansions of Feynman diagrams}},  in \emph{{6th International Workshop on
  New Computing Techniques in Physics Research: Software Engineering,
  Artificial Intelligence Neural Nets, Genetic Algorithms, Symbolic Algebra,
  Automatic Calculation (AIHENP 99) Heraklion, Crete, Greece, April 12-16,
  1999}}, 1999 [\href{https://arxiv.org/abs/hep-ph/9905298}{{\ttfamily
  hep-ph/9905298}}].

\bibitem{Harlander:1999cs}
R.~Harlander, \emph{{Asymptotic expansions: Methods and applications}},
  {\emph{Acta Phys. Polon.} {\bfseries B30} (1999) 3443}
  [\href{https://arxiv.org/abs/hep-ph/9910496}{{\ttfamily hep-ph/9910496}}].

\bibitem{Dlapa:2020cwj}
C.~Dlapa, J.~Henn and K.~Yan, \emph{{Deriving canonical differential equations
  for Feynman integrals from a single uniform weight integral}},
  \href{https://doi.org/10.1007/JHEP05(2020)025}{\emph{JHEP} {\bfseries 05}
  (2020) 025} [\href{https://arxiv.org/abs/2002.02340}{{\ttfamily
  2002.02340}}].

\bibitem{Henn:2020lye}
J.~Henn, B.~Mistlberger, V.A.~Smirnov and P.~Wasser, \emph{{Constructing d-log
  integrands and computing master integrals for three-loop four-particle
  scattering}}, \href{https://doi.org/10.1007/JHEP04(2020)167}{\emph{JHEP}
  {\bfseries 04} (2020) 167}
  [\href{https://arxiv.org/abs/2002.09492}{{\ttfamily 2002.09492}}].

\bibitem{Lee:2011jt}
R.N.~Lee, A.V.~Smirnov and V.A.~Smirnov, \emph{{Master Integrals for Four-Loop
  Massless Propagators up to Transcendentality Weight Twelve}},
  \href{https://doi.org/10.1016/j.nuclphysb.2011.11.005}{\emph{Nucl. Phys.}
  {\bfseries B856} (2012) 95}
  [\href{https://arxiv.org/abs/1108.0732}{{\ttfamily 1108.0732}}].

\bibitem{Panzer:2014caa}
E.~Panzer, \emph{{Algorithms for the symbolic integration of hyperlogarithms
  with applications to Feynman integrals}},
  \href{https://doi.org/10.1016/j.cpc.2014.10.019}{\emph{Comput. Phys. Commun.}
  {\bfseries 188} (2015) 148}
  [\href{https://arxiv.org/abs/1403.3385}{{\ttfamily 1403.3385}}].

\bibitem{Borowka:2017idc}
S.~Borowka, G.~Heinrich, S.~Jahn, S.P.~Jones, M.~Kerner, J.~Schlenk et~al.,
  \emph{{pySecDec: a toolbox for the numerical evaluation of multi-scale
  integrals}}, \href{https://doi.org/10.1016/j.cpc.2017.09.015}{\emph{Comput.
  Phys. Commun.} {\bfseries 222} (2018) 313}
  [\href{https://arxiv.org/abs/1703.09692}{{\ttfamily 1703.09692}}].

\bibitem{Kniehl:2017ikj}
B.A.~Kniehl, A.F.~Pikelner and O.L.~Veretin, \emph{{Three-loop massive tadpoles
  and polylogarithms through weight six}},
  \href{https://doi.org/10.1007/JHEP08(2017)024}{\emph{JHEP} {\bfseries 08}
  (2017) 024} [\href{https://arxiv.org/abs/1705.05136}{{\ttfamily
  1705.05136}}].

\end{thebibliography}\endgroup

\end{document}